\documentclass[prd,10pt,twocolumn,tightenlines,superscriptaddress,numerical,notitlepage,showpacs,amsmath,amssymb,amsfonts,aps,nofootinbib,longbibliography,floatfix]{revtex4-1}
\usepackage[utf8]{inputenc}
\usepackage[justification=raggedright,singlelinecheck=false,font=small]{caption}
\usepackage{graphicx,bm,bbm,color,amssymb,amsmath}
\usepackage{multirow}
\usepackage{subcaption}
\usepackage{slashed,verbatim}
\usepackage{orcidlink}
\usepackage{soul}
\usepackage[usenames,dvipsnames]{xcolor}
\sethlcolor{red}
\usepackage{ulem}
\usepackage{combelow}
\usepackage{dsfont}
\usepackage{amsmath,amssymb,mathtools,bm,bbm,braket,slashed}
\usepackage{tikz}
\usepackage{cancel}
\usepackage{xcolor}
\usepackage{ulem}
\usepackage{hyperref}
\hypersetup{colorlinks=true, citecolor=blue, urlcolor=blue, citecolor=red}

\definecolor{purple}{rgb}{0.8,0,0.6}

\definecolor{battleshipgrey}{rgb}{0.2, 0.52, 0.51}
\definecolor{darkgreen}{rgb}{0.12, 0.5, 0.17}

\renewcommand\sout{\bgroup\color{blue} \ULdepth=-.5ex \ULset}
    \newcommand{\beqn}{\begin{eqnarray}}
    \newcommand{\eeqn}{\end{eqnarray}}
    \newcommand{\beqs}{\begin{subequations}}
    \newcommand{\eeqs}{\end{subequations}\\[-2mm]\noindent}

\begin{document}
%\title{Effect of background magnetic field on virtual photon bremsstrahlung in quark-gluon-plasma}
\title{Virtual photon bremsstrahlung in a magnetized quark-gluon-plasma}

\author{Aritra Bandyopadhyay\orcidlink{0000-0002-3058-7258}}
\email{aritrabanerjee.444@gmail.com }
\affiliation{Institute of Theoretical Physics, University of Wrocław, plac Maksa Borna 9, PL-50204 Wrocław, Poland}

\author{Krzysztof Redlich\orcidlink{0000-0002-2629-1710}}
\email{krzysztof.redlich@uwr.edu.pl}
\affiliation{Institute of Theoretical Physics, University of Wrocław, plac Maksa Borna 9, PL-50204 Wrocław, Poland}
\affiliation{Polish Academy of Sciences PAN, Podwale 75, PL-50449 Wroclaw, Poland}

\author{Chihiro Sasaki\orcidlink{0000-0003-4612-3375}}
\email{chihiro.sasaki@uwr.edu.pl}
\affiliation{Institute of Theoretical Physics, University of Wrocław, plac Maksa Borna 9, PL-50204 Wrocław, Poland}
\affiliation{International Institute for Sustainability with Knotted Chiral Meta Matter (WPI-SKCM$^2$), Hiroshima University, 1-3-1 Kagamiyama, 739-8531, Higashi-Hiroshima, Hiroshima, Japan}

\date{\today}

\begin{abstract}
We investigate dilepton production via virtual quark bremsstrahlung in a hot, magnetized quark-gluon plasma (QGP), extending the Landau-Pomeranchuk-Migdal (LPM) framework to account for a background magnetic field and for Fermi-Dirac quark statistics. An external electromagnetic field drives the quark distribution out of equilibrium, and this off-equilibrium correction, obtained from the Boltzmann equation in the relaxation-time approximation, is in principle explicit in the electric and magnetic field strengths. We argue that this correction, being linear in the electric field, vanishes identically upon integration over the full dilepton solid angle and therefore does not contribute to the dilepton production rate (DPR). The leading magnetic-field signature in the rate instead enters entirely through the equilibrium sector: the medium density and Debye screening mass, computed from the Landau-level structure of the magnetized QGP, modifying the LPM suppression parameter. Working in the regime relevant to the hydrodynamic phase of heavy-ion collisions, we present the LPM-induced DPR for quark scatterers and establish, for the first time, a nonzero magnetic-field imprint on this observable through the LPM mechanism.
\end{abstract}

\maketitle

\section{Introduction}

Electromagnetic (EM) probes, such as dileptons and photons are considered to be among the most important probes for exploring hot and dense quark matter~\cite{Shuryak:1978ij,McLerran:1984ay,Kajantie:1986dh}. Such claims about EM probes are aided by their two primary features, i.e., less contamination because they have a large mean free path being external to the strongly interacting bulk medium; and carrying maximum information because of being produced from all the important stages of a heavy ion collision (HIC). Separating the contributions to the dilepton/photon production from different stages of HIC is essential to have information regarding that particular stage. Dileptons have a distinct advantage in this aspect, since they are associated with an additional kinematic variable, their invariant mass $M$. Being Lorentz invariant, $M$ is unaffected by the collective flow of the medium, and different ranges of $M$ are dominated by different production mechanisms and different stages of the collision~\cite{Rapp:2014hha}. The most basic ingredient to obtain the dilepton invariant mass spectra is the dilepton production rate (DPR) which is the central theme of the present study.

Recent measurements at SPS~\cite{NA60:2006ymb}, RHIC~\cite{STAR:2013pwb,PHENIX:2015vek,STAR2022,STAR2023b} and the LHC~\cite{ALICE:2023jef} have renewed interest in precise theoretical rates, especially in the regions of low and intermediate invariant mass. In a quark-gluon plasma (QGP), low invariant mass lepton pairs originate predominantly from virtual bremsstrahlung of quarks scattering in the medium~\cite{Haglin:1992ew}. This process is strongly affected by the Landau-Pomeranchuk-Migdal (LPM) effect~\cite{Cleymans:1992je,LandauPomeranchuk1953a,LandauPomeranchuk1953b,Migdal:1956tc}: when the formation time of the radiated photon exceeds the time between successive collisions, emission amplitudes from different segments of the quark's path interfere destructively, and the rate is suppressed relative to the incoherent sum.

The LPM effect was applied to DPR, first to pion bremsstrahlung in a hadronic medium~\cite{Cleymans1993had} and then to quark bremsstrahlung in the QGP~\cite{Cleymans:1992je}. In the QGP case, it was shown \cite{Cleymans:1992je}  that multiple scattering renders the rate infrared stable for massless quarks, and that the LPM suppression of the low invariant mass DPR outweighs the effect of the thermal quark mass. In this work, we extend the previous studies of virtual-photon bremsstrahlung~\cite{Cleymans:1992je} to a magnetized QGP employing Fermi-Dirac rather than Boltzmann statistics and study the resulting low invariant mass DPR.

The magnetized QGP has attracted sustained attention over the past two decades because of the geometry of non-central collisions. The spectator nuclei generate a magnetic field of order $eB\sim(1$--$10)\,m_\pi^2$ at early times, among the strongest in the present-day universe~\cite{Kharzeev:2007,Skokov:2009,Deng:2012,Huang:2016}. This field decays quickly in the pre-equilibrium stage~\cite{McLerran:2014,Siddique:2025tzd}. $\Lambda$-hyperon polarization measurements point to a weak field surviving into the late stages~\cite{STAR2023a}, leaving a QGP in a weak background field, with $eB$ small compared to the hard scales of the process. EM probes couple directly to such a field, making them natural candidates for a QGP magnetometer. The DPR in a magnetized medium has been studied extensively within field-theoretic approaches, using the Schwinger proper-time propagator in both its limiting and general forms~\cite{Mustafa:2025uad,Bandyopadhyay:2016fyd,Das:2021fma,Ghosh:2018xhh,Wang:2022jxx,Wang:2023fst} and the Ritus eigenfunction method~\cite{Sadooghi:2016jyf}. Kinetic-theory studies have instead focused on the non-equilibrium correction to the DPR induced by the electromagnetic field~\cite{Sun2022,Sun2024a,Sun2024b,Gao2025}. A recent hydrodynamic study~\cite{Gao2025} found this field-driven leading order (linear in the field) correction to leave essentially no imprint on the invariant-mass spectrum, while strongly affecting the dilepton elliptic flow. We argue that this absence follows from symmetry, so that the leading magnetic-field dependence of the rate must arise from the equilibrium medium.

In a QGP elastic scattering is governed by $t$-channel gluon exchange. Small-angle deflections therefore dominate, and successive collisions of a quark are correlated rather than independent~\cite{Cleymans:1992je}. The same situation arises for fast charged particles traversing solids~\cite{Migdal:1956tc,Kalashnikov1985}, where the kinetic equation for the trajectory correlations reduces, in the small-angle limit, to a Fokker--Planck equation. In this description, instead of the elastic cross section, the mean-square scattering angle per unit path length becomes the primary parameter (denoted as $q$) of the system~\cite{Cleymans:1992je}. For Coulomb-like scattering the angular integral is logarithmically sensitive to small angles, and in the plasma this sensitivity is cut off by Debye screening. As a result, $q$ depends on the medium through the density of scatterers $n$ and the Debye screening mass $m_D$, and it sets the strength of the LPM suppression. This structure makes $q$ the natural entry point for a background magnetic field. Landau quantization reorganizes the transverse phase space of the charged quarks, modifying both $n(T,eB)$ and $m_D(T,eB)$, and thereby the degree of coherent suppression of low-mass bremsstrahlung. To the best of our knowledge, this magnetic modification of the LPM parameter has not been studied previously.

The paper is organized as follows. In Sec.~\ref{sec:formalism} we first review the formalism for virtual-photon bremsstrahlung with Boltzmann statistics (Sec.~\ref{subsec:thermal_recap}). We then introduce Fermi--Dirac statistics and the magnetic-field corrections through the Landau-quantized medium quantities, and show that the field-induced non-equilibrium correction to the quark distribution does not contribute to the DPR $dN/d^4xdM$ (Sec.~\ref{subsec:magnetic_corrections}). Our numerical results are presented in Sec.~\ref{sec:results}, and Sec.~\ref{sec:conclusions} summarizes our findings and gives an outlook.

%% ============================================================
\section{Formalism}
\label{sec:formalism}
%% ============================================================

%% ============================================================
\subsection{Thermal Dilepton Rate with the Landau-Pomeranchuk Effect}
\label{subsec:thermal_recap}
%% ============================================================

In this subsection we briefly recapitulate the calculation of the dilepton production rate from virtual quark bremsstrahlung in a thermal QGP, as derived in Ref.~\cite{Cleymans:1992je}. We follow the notation of that reference closely, and use this as the foundation upon which the magnetic-field corrections will be built in the following section.

The physical process under consideration is virtual bremsstrahlung of quarks moving through the QGP: a fast quark scatters off the color charges in the medium and radiates a virtual photon $\gamma^*$, which subsequently decays into a lepton pair $\ell^+\ell^-$. At low invariant masses $M$, this process dominates over quark-antiquark annihilation as the primary source of dileptons in the QGP~\cite{Cleymans:1992je}.

The dilepton production rate is related to the rate of
virtual photon production through the standard formula~\cite{Kajantie:1986dh,Cleymans:1986na}:
\begin{equation}
\frac{dN^{l^+l^-}}{d^4x\, dM} = \frac{2\alpha}{3\pi M}
\left(1 - \frac{4m_\ell^2}{M^2}\right)^{1/2}
\left(1 + \frac{2m_\ell^2}{M^2}\right)
\frac{dN^{\gamma^*}}{d^4x},
\label{eq:dilepton_from_vp}
\end{equation}
where $\alpha$ is the fine structure constant, $M$ is the dilepton invariant mass, and $m_\ell$ is the lepton mass. This expression is purely kinematic: it contains no medium physics and simply converts a virtual photon flux into a dilepton yield. For dielectron production, the threshold factors $(1 - 4m_\ell^2/M^2)^{1/2}$ and $(1 + 2m_\ell^2/M^2)$ are very close to unity and are neglected in what follows~\cite{Cleymans:1992je}.

To obtain the total virtual photon rate $dN^{\gamma^*}/d^4x$, one integrates the single-quark virtual photon emission rate $dN^{\gamma^*}/d\omega\,dt$ over the quark phase space, weighted by the quark distribution function. Following~\cite{Cleymans:1992je}, averaging over all emitting quarks and antiquarks in a unit volume gives:
\begin{multline}
\frac{dN^{\gamma^*}}{d^4x} = \Sigma_f\, 2d_q
\int d^3p_1\, f(\mathbf{p}_1) \\
\times \frac{\int d^3p_2\, f(\mathbf{p}_2)\, |\mathbf{v}_{12}| \int_M^{\Delta} d\omega\, \frac{dN^{\gamma^*}}{d\omega\, dt}}{\int d^3p_2\, f(\mathbf{p}_2)\, |\mathbf{v}_{12}|},
\label{eq:vp_phase_space}
\end{multline}
where $f(\mathbf{p})$ is the quark momentum distribution function, $|\mathbf{v}_{12}| = [s(s-4m_q^2)]^{1/2}/2E_1E_2$
is the relative velocity between the emitting quark and a scatterer with $s = 2(m_q^2 + E_1 E_2 - \mathbf{p}_1\cdot\mathbf{p}_2)$, $d_q = 2_s \times 3_c = 6$ is the quark spin-color degeneracy factor, the overall factor of 2 accounts for antiquarks, and $\Delta$ is the kinematic upper limit on the radiated photon energy imposed by energy-momentum conservation.

The single-quark emission rate $dN^{\gamma^*}/d\omega\,dt$ encodes the LPM physics. Starting from the soft-photon approximation for the energy radiated per unit photon momentum by a charged particle moving along a trajectory $\mathbf{r}(t)$~\cite{Cleymans:1992je,Migdal:1956tc}, and averaging over all possible trajectories in the medium using the Fokker-Planck equation for the density matrix~\cite{Migdal:1956tc}, one arrives at the following result for the number of virtual photons emitted per unit frequency interval and per unit time by a fast quark moving in a QGP medium~\cite{Cleymans:1992je}:
\begin{equation}
\frac{dN^{\gamma^*}}{d\omega\, dt} = \frac{8e_q^2\alpha k}{3\pi\omega^2}q
\left[\left(\frac{m_q^2}{p_1^2} + \frac{M^2}{\omega^2}\right)^2
+ \frac{16q}{9\omega}\right]^{-1/2}.
\label{eq:vp_rate_221}
\end{equation}
where $k = |\mathbf{k}|$ is the photon three-momentum, $\omega$ is the photon energy, $M^2 = \omega^2 - k^2$ is the photon invariant mass squared, $e_q$ is the quark electric charge, and $p_1$ is the quark momentum. The LPM parameter $q$ appearing in Eq.~(\ref{eq:vp_rate_221}) is the mean squared transverse momentum transfer per unit path length:
\begin{equation}
q \equiv \frac{\langle\theta_S^2\rangle}{4}
= \frac{1}{4}\left\langle n\, v_{12}
\int d\Omega\, \theta^2\frac{d\sigma}{d\Omega}\right\rangle,
\label{eq:q_def}
\end{equation}
where $n$ is the density of scatterers in the medium, $\theta$ is the scattering angle, $d\sigma/d\Omega$ is the elastic differential cross section, and the average is over the thermal distribution. The parameter $q$ quantifies how effectively the medium destroys the coherence of the radiated virtual photon: it is large when scatterings are frequent and hard, leading to strong LPM suppression, and small when scatterings are rare and soft, leaving the radiation closer to the Bethe-Heitler limit.

To evaluate $q$ explicitly in the thermal QGP, one inserts the small-angle elastic cross section for $t$-channel gluon exchange~\cite{Cleymans:1992je}:
\begin{equation}
\frac{d\sigma_{ab\to ab}}{dt} = \frac{2\pi\alpha_s^2 C_{ab}}{t^2},
\quad C_{ab} = 4/9,
\label{eq:elastic_xsec}
\end{equation}
where $t$ is the Mandelstam variable and we have retained only the $qq\to qq$ channel, since only the quark density $n(T,eB)$ and screening mass $m_D(T,eB)$ are modified by $B$ in our treatment. Using the kinematic relation $d\sigma/d\Omega = (p_1^2/\pi)\,d\sigma/dt$ and $t \approx -p_1^2\theta^2$ in the small-angle limit, the angular integral in Eq.~(\ref{eq:q_def}) becomes:
\begin{equation}
\int d\Omega\,\theta^2\frac{d\sigma}{d\Omega}
= \frac{4\pi\alpha_s^2 C_{ab}}{p_1^2}
\int_{\theta_{\min}}^{\theta_{\max}}\frac{d\theta}{\theta}
= \frac{4\pi\alpha_s^2 C_{ab}}{p_1^2}\,L_C,
\label{eq:angular_integral}
\end{equation}
where the Coulomb logarithm $L_C = \ln(\theta_{\max}/\theta_{\min})$ arises from the logarithmic divergence of the $t$-channel integral, regulated in the infrared by the Debye screening mass $m_D$ through $\theta_{\min} \sim m_D/p_1$, giving $L_C = \ln(p_1/m_D)$.

The thermal average $\langle n\,v_{12}\,(\cdots)\rangle$ is evaluated by integrating over the scatterer momentum $p_2$ with the Boltzmann distribution $f(p_2) = (2\pi)^{-3}e^{-p_2/T}$:
\begin{equation}
n^{\rm BM}(T) = \int\frac{d^3p_2}{(2\pi)^3}e^{-p_2/T}= \frac{T^3}{\pi^2},
\label{eq:thermal_density}
\end{equation}
where we used $\int_0^\infty x^2 e^{-x}dx = 2$, and the relative velocity $v_{12}\to 1$ in the massless ultrarelativistic limit. Summing over the $qq$ channels with the appropriate color-spin-flavour degeneracy factors gives $d = \frac{4}{9}\times 2\times 2_s\times 2_f\times 3_c = 10\frac{2}{3}$. 

Collecting all factors, one obtains the standard thermal result~\cite{Cleymans:1992je}:
\begin{equation}
q^{\rm BM}(T,p_1) = C\,L_C\frac{\alpha_s^2\,T^3}{p_1^2} \, ,
\label{eq:q0_thermal}
\end{equation}
where $C = d/\pi \approx 3.4$ collects the degeneracy factors.

%% ============================================================
\subsection{Magnetic-Field Corrections through Equilibrium Static Quantities}
\label{subsec:magnetic_corrections}
%% ============================================================

We now turn to the leading effect of a background magnetic field on the dilepton invariant-mass spectrum. In principle, an external electromagnetic field drives the quark distribution away from equilibrium, $f = f_0 + \delta f$, and this off-equilibrium correction $\delta f$ could in turn modify the dilepton rate. However, $\delta f$ is linear in the fluid-frame electric field $\mathbf{E}$, so at linear order the only scalar that can be formed at fixed pair momentum $\mathbf{K}$ is $\mathbf{K}\cdot\mathbf{E}$. The equilibrium weight multiplying it is even under $\mathbf{K}\to-\mathbf{K}$, since a thermal medium with a background $\mathbf{B}$ is parity invariant, whereas $\mathbf{K}\cdot\mathbf{E}$ is odd. The invariant-mass spectrum is obtained by integrating over the full solid angle of $\mathbf{K}$, so this term integrates to zero identically. Consequently, the off-equilibrium sector does not contribute to $dN/d^4xdM$ at all: this is a direct consequence of $dN/d^4xdM$ being a rotational scalar with no surviving vector at linear order in $\mathbf{E}$, and is corroborated by the numerical results of Ref.~\cite{Gao2025}, where the analogous off-equilibrium correction was found to leave $dN/dM$ unchanged while producing a sizable effect on the elliptic flow $v_2(M)$ (a projection onto the $\cos 2\phi_K$ harmonic, which is not subject to the same cancellation). The entire leading magnetic-field imprint on $dN/d^4xdM$ therefore originates instead from the equilibrium medium quantities $n(T,eB)$ and $m_D(T,eB)$ that enter the LPM suppression parameter $q$, which we now compute.

In the presence of a uniform magnetic field $\mathbf{B} = B\hat{z}$, the quark transverse momenta are quantized into discrete Landau levels. The continuous phase space integral is replaced by:
\begin{equation}
\int\frac{d^3p}{(2\pi)^3} \longrightarrow\sum_{k=0}^{\infty}\alpha_k\frac{|q_f|B}{4\pi}\int_{-\infty}^{+\infty}\frac{dp_z}{2\pi},\label{eq:phase_space_replacement}
\end{equation}
where $\alpha_k = 2 - \delta_{k,0}$ is the spin degeneracy and the quark energy in the $k$-th Landau level is:
\begin{equation}
\omega_{f,k} = \left(p_z^2 + m_q^2 + 2k|q_f|B\right)^{1/2}.
\label{eq:LL_energy}
\end{equation}
It is from this replacement alone that the magnetic field modifies the equilibrium properties of the medium, and hence provides the sole channel through which $B$ enters the DPR within the present approximation. The medium density in the magnetized QGP, per color, is:
\begin{equation}
n(T,eB) = 2\sum_{f=u,d}\sum_{k=0}^{\infty}\alpha_k\frac{|q_f|B}{4\pi^2}\int_{-\infty}^{+\infty}dp_z\,f_0\!\left(\omega_{f,k}\right),
\label{eq:density_total}
\end{equation}
where the factor of 2 accounts for quarks and antiquarks, and $f_0(\omega_{f,k}) = 1/(e^{\omega_{f,k}/T}+1)$ is the Fermi-Dirac distribution. The Debye screening mass is defined through the static limit of the longitudinal gluon self-energy, $m_D^2 = -\Pi^{00}(0,\mathbf{q}\to 0)$. Applying the replacement~(\ref{eq:phase_space_replacement}) to the one-loop quark contribution gives:
\begin{multline}
m_D^2(T,eB) = \frac{g^2}{4\pi^2}\sum_{f=u,d}\sum_{k=0}^{\infty}\alpha_k|q_f|B \\
\int_{-\infty}^{+\infty}dp_z\,\beta f_0(\omega_{f,k})[1-f_0(\omega_{f,k})].\label{eq:mD_LL}
\end{multline}
The gluon contribution to $m_D$ is unaffected by $B$ since gluons carry no electric charge, and it is omitted here, consistently with the quark-scatterer choice.

Substituting into Eq.~(\ref{eq:q0_thermal}), the $B$-dependent LPM parameter is:
\begin{equation}
    q(T,eB,p_1) = \frac{3\zeta(3)}{4}\,C\,\frac{\alpha_s^2\,T^3}{p_1^2}\;
\frac{n(T,eB)}{n_0(T)}\;
\ln\!\left(\frac{p_1}{m_D(T,eB)}\right),
\label{eq:q_magnetised}
\end{equation}
Here $n_0(T)$, and $m_{D0}(T)=gT\sqrt{N_f/6}$ denote the field-free values, which give $q_0(T,p_1)=\tfrac{3\zeta(3)}{4}\,C\,\alpha_s^2T^3\ln(p_1/m_{D0})/p_1^2$, the Fermi--Dirac counterpart of $q^{\rm BM}$, Eq.~\eqref{eq:q0_thermal}, with $L_C\to\ln(p_1/m_{D0})$. The factor $3\zeta(3)/4$ is the Fermi--Dirac to Boltzmann ratio of the scatterer density per degree of freedom; only the ratio $n/n_0$ enters $q$, so the overall normalization of $n$ in Eq.~\eqref{eq:density_total} is immaterial.

$q(T,eB,p_1)$ is evaluated numerically by truncating the Landau sum at sufficiently high $k$ to ensure convergence. Because Eqs.~(\ref{eq:density_total}) and (\ref{eq:mD_LL}) are exact Landau-level sums rather than a truncated weak-field expansion, they remain valid for arbitrary $eB/T^2$; convergence of the truncated sum is a numerical, not a physical, requirement, and no additional hierarchy between $eB$ and $T^2$ is needed for these equilibrium quantities.

\par
The lower limit of the $p_1$ integration is set to $p_1 = m_D$, since the small-angle approximation underlying the LPM framework is valid only for $p_1 \geq m_D$. Below this scale, the test quark momentum falls within the screening cloud of the medium, and it can no longer be treated as a hard particle undergoing perturbative small-angle deflections. Consequently, the Coulomb logarithm $\ln(p_1/m_D)$, which represents the ratio $\ln(\theta_{\rm max}/\theta_{\rm min})$ of the maximum to minimum scattering angles, becomes negative for $p_1 < m_D$, rendering the LPM suppression parameter $q$ unphysical. Imposing $p_1 \geq m_D$ as the lower integration limit ensures that $\ln(p_1/m_D) > 0$ throughout the integration domain, keeping $q$ positive definite and the small-angle approximation self-consistent. In the magnetized case, the same argument applies with $m_D$ replaced by the field-dependent Debye mass $m_D(T, eB)$, computed from the Landau-level quantized quark propagator.

The framework rests on a single hierarchy of scales, $eB \lesssim M^2$, which justifies retaining the field-free phase space $d^3p/(2\pi)^3$ with dispersion $\omega = (\mathbf{p}^2+m_q^2)^{1/2}$ in the dilepton production vertex of Eq.~(\ref{eq:vp_phase_space}), rather than the Landau-quantized form of Eq.~(\ref{eq:phase_space_replacement}). Quantitatively, the ratio of the LPM coherence length $l_c \sim p_1/M^2$ to the Larmor radius $R_L \sim T/|q_f|B$ is
\begin{equation}
\frac{l_c}{R_L} \sim \frac{eB}{M^2} \lesssim 1,
\label{eq:scale_ratio}
\end{equation}
so the quark completes only a small fraction of a cyclotron orbit over the formation length, and its propagation between scatterings is well described as a straight-line random walk in transverse momentum space -- the basis of the Migdal approach. For the weak fields expected in the late, hydrodynamic stage, $eB\lesssim m_\pi^2\approx0.02~\mathrm{GeV}^2$, this condition is well satisfied already for $M\gtrsim0.3$~GeV ($eB/M^2\lesssim0.25$, see Table~\ref{tab:scale-check}), which validates the mixed treatment adopted here: field-free phase space for the dilepton production vertex, and Landau-quantized equilibrium quantities for the medium response entering the LPM parameter. We also show results up to $eB=0.4~\mathrm{GeV}^2\approx20\,m_\pi^2$, i.e.\ at and above the upper end of the early-time estimates quoted in the Introduction. These fields are exploratory: the hierarchy holds there only at the upper end of the $M$ range, and the corresponding curves indicate the trend (Sec.~\ref{subsec:dilepton-rates}). Throughout, we work to linear order in $\mathbf{E}$. Nonlinear $\mathcal{O}(E^2)$ corrections to $\delta f$ survive the angular integration and, for $E\sim B$, are of the same order as the $\mathcal{O}((eB)^2)$ equilibrium effect computed here; they lie beyond the accuracy of the present calculation and are left to a future, second-order treatment.

\section{Results}
\label{sec:results}

\begin{figure*}[tbh]
    \centering
    \includegraphics[width=\linewidth]{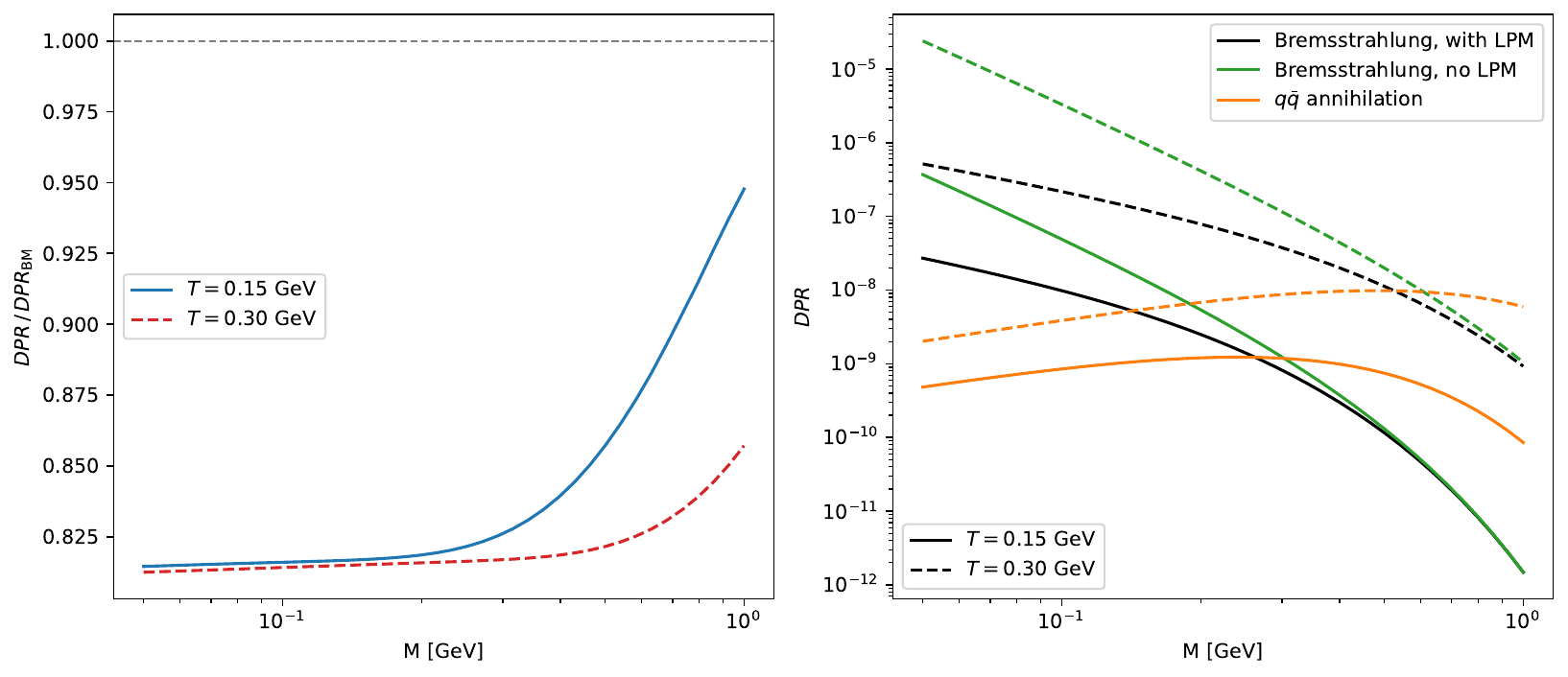}
    \caption{Left: Ratio of the dilepton production rate computed with the Fermi-Dirac quark distribution to that computed with the Boltzmann distribution, as a function of $M$, at $T=0.15$ and $0.30$~GeV. Right: Fermi-Dirac bremsstrahlung rate with LPM suppression, the same rate without LPM, and $q\bar q$ annihilation, at the same two temperatures (solid: $T=0.15$~GeV, dashed: $T=0.30$~GeV). Both panels are for $eB=0$, and within the constant-$L_C$ approximation ($L_C=1$).}
    \label{fig:bm-fd-comparison}
\end{figure*}

Unless stated otherwise we use $N_c =3$, $N_f=2 (u,d)$, and $m_q=5$ MeV. The Debye mass uses the running coupling $g^2(T) = 24\pi^2/[(11N_c-2N_f)\ln(2\pi T/\Lambda)]$ with $\Lambda=176$ MeV. The lower cutoff is $p_1\geq m_D$, and the two temperatures are $T=0.15$ and $0.30$ GeV. Within the figures DPR denotes the rate per unit space-time volume, $\mathrm{DPR}=dN/d^4x\,dM$, of a uniform medium, evaluated with Fermi-Dirac statistics unless labelled $\mathrm{DPR}_{\rm BM}$ (Boltzmann).

\subsection{Numerical convergence of the Landau-level truncation}
\label{subsec:kmax-convergence}

The Landau sums defining $n(T,eB)$ and $m_D^2(T,eB)$ in Eqs.~\eqref{eq:density_total} and~\eqref{eq:mD_LL} run formally to $k=\infty$; in practice they must be truncated at a finite $k_{\max}$. The level spacing in transverse momentum, $\Delta(p_\perp^2) = 2|q_f|B$, depends linearly on $B$, and hence a $k_{\max}$ adequate at strong field can substantially under-count $n(T,eB)$ and $m_D(T,eB)$ at weak field, and only in the limit $k_{\max}\to\infty$ does the Landau sum reduce to the continuum result $n_0(T)$ (Sec.~\ref{subsec:magnetic_corrections}) as $eB\to0$.

We fix $k_{\max}$ systematically for each $\{T,eB,q_f\}$ by doubling it until the resulting $n(T,eB)$ changes by less than $0.1\%$:
\begin{equation}
\left|\frac{n(T,eB;2k_{\max}) - n(T,eB;k_{\max})}{n(T,eB;2k_{\max})}\right|< 10^{-3}\,,
\label{eq:convergence-criterion}
\end{equation}
starting from $k_{\max}=10$. The same $k_{\max}$ is applied to $m_D(T,eB)$, taking the larger of the values required by the $u$- and $d$-quark sums individually (their charges $q_u=2/3$, $q_d=1/3$ give different level spacings and hence different convergence rates at fixed $B$). As a cross-check, we verified that with $k_{\max}$ fixed by Eq.~\eqref{eq:convergence-criterion}, $n(T,eB)/n_0(T) \to 1$ and $m_D(T,eB)/m_{D0}(T) \to 1$ to within $10^{-4}$ in the $eB\to0$ limit, confirming that the Landau-quantized medium response correctly reduces to the field-free Fermi-Dirac continuum result.

Table~\ref{tab:kmax-convergence} lists the resulting $k_{\max}$ for the field strengths and temperatures used in this work. As expected, $k_{\max}$ scales approximately as $1/eB$ at fixed $T$, and grows with $T$ at fixed $eB$. For the weakest field considered here, $eB=0.001~\mathrm{GeV}^2$, several thousand to several tens of thousands of Landau levels are required for convergence.

\begin{table}[h]
\centering
\caption{Landau level cutoff $k_{\max}$ required for $n(T,eB)$ to satisfy the convergence criterion of Eq.~\eqref{eq:convergence-criterion}, for the field strengths and temperatures used in this work.}
\label{tab:kmax-convergence}
\begin{tabular}{cccc}
\hline\hline
$T$ [GeV] & $eB$ [GeV$^2$] & $k_{\max}$ & $n(T,eB)/n_0(T)$ \\
\hline
\multirow{5}{*}{0.15}
 & 0.001 & 10240 & 1.000 \\
 & 0.02  & 640   & 1.006 \\
 & 0.2   & 40    & 1.349 \\
 & 0.3   & 40    & 1.651 \\
 & 0.4   & 20    & 1.991 \\
\hline
\multirow{5}{*}{0.30}
 & 0.001 & 40960 & 1.000 \\
 & 0.02  & 2560  & 1.000 \\
 & 0.2   & 160   & 1.034 \\
 & 0.3   & 160   & 1.069 \\
 & 0.4   & 80    & 1.113 \\
\hline\hline
\end{tabular}
\end{table}

We emphasize that a fixed, field-independent truncation (e.g. $k_{\max}=20$) is adequate only for the strongest fields in Table~\ref{tab:kmax-convergence}; applied uniformly across the full range it under-converges $n(T,eB)$ and $m_D(T,eB)$ at weak field by up to an order of magnitude (e.g. at $eB=0.001$ GeV$^2$ it gives $n/n_{\rm conv} \approx 0.15$ at $T=0.15$ GeV and $\approx 0.04$ at $T=0.30$ GeV), which can distort the relative ordering of the dilepton rate curves as a function of $eB$. All results shown in this work use the converged, field-dependent $k_{\max}$ of Table~\ref{tab:kmax-convergence} rather than a fixed value.

\begin{figure*}[tbh]
    \centering
    \includegraphics[width=\linewidth]{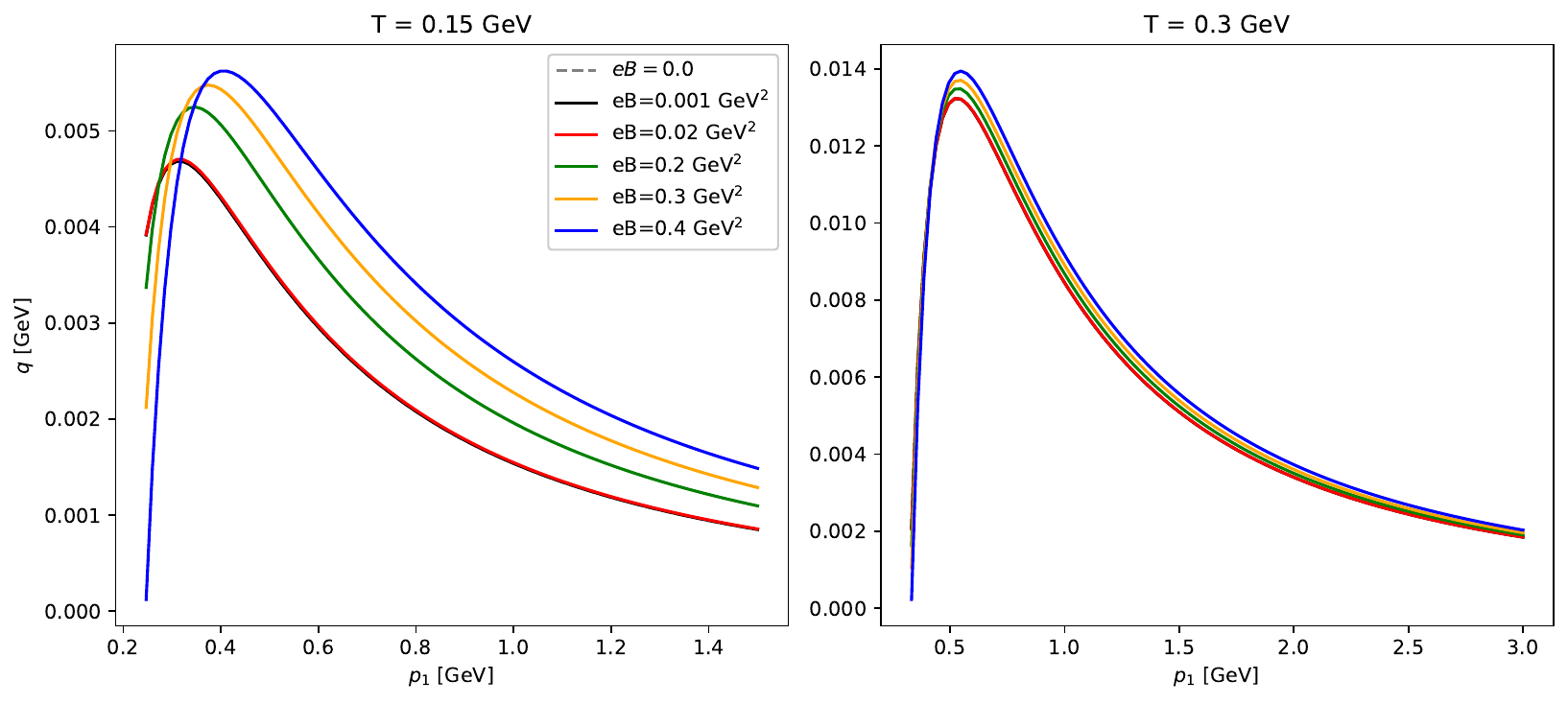}
    \caption{Variation of the LPM parameter $q(T,eB,p_1)$ with the quark momentum $p_1$, for different values of the external magnetic field and two fixed temperatures, $T=0.15$~GeV (left) and $T=0.30$~GeV (right). The dashed curve is the field-free result, i.e.\ Eq.~\eqref{eq:q_magnetised} at $eB=0$.}
    \label{fig:q-vs-p1}
\end{figure*}

\subsection{Boltzmann versus Fermi-Dirac statistics}
\label{subsec:bm-fd-comparison}

Before presenting the magnetized-field results, we directly quantify the impact of using the Fermi-Dirac distribution instead of the Boltzmann distribution, which was used in Ref.~\cite{Cleymans:1992je} to obtain complete analytical expressions. The left panel of Fig.~\ref{fig:bm-fd-comparison} shows the ratio of $dN/d^4x\,dM$ evaluated with the full Fermi-Dirac $f_0$ to the same quantity evaluated with a Boltzmann quark distribution, $f_0\to e^{-p/T}$, at fixed $T$ and $eB=0$. In both cases the LPM parameter ($q^{\rm BM}$ and $q_0$, respectively) is evaluated with the constant Coulomb-logarithm approximation of Ref.~\cite{Cleymans:1992je} ($L_C=1$ in Eq.~\eqref{eq:q0_thermal}, and $\ln(p_1/m_{D0})\to1$ in the Fermi-Dirac case), so that the comparison isolates the effect of quantum statistics alone, independent of the $\ln(p_1/m_D)$ treatment discussed elsewhere in this work.

The Fermi-Dirac rate lies systematically below the Boltzmann rate at both temperatures. The ratio $\mathrm{DPR}/\mathrm{DPR}_{\rm BM}$ starts at $\approx0.82$ ($T=0.15$~GeV) and $\approx0.81$ ($T=0.30$~GeV) at the smallest invariant mass considered ($M=0.05$~GeV), and rises monotonically to $\approx0.95$ and $\approx0.86$, respectively, at $M=1$~GeV. This $5$--$20\%$ suppression arises because the Fermi-Dirac occupation number, $f_0=e^{-p/T}/(1+e^{-p/T})$, is smaller than the Boltzmann one (a statistics effect, not Pauli blocking), most strongly for $p\lesssim T$. Low-mass pairs probe this region with larger weight, and so does a higher temperature at fixed $M$, whereas for $p\gg T$ the two distributions coincide and the ratio tends to unity. The Fermi-Dirac normalization of the scatterer density entering $q$, $3\zeta(3)/4\approx0.90$ times the Boltzmann value, acts in the opposite direction (it weakens the LPM suppression and raises the Fermi-Dirac rate by up to $\approx5\%$), so the net suppression is smaller than that from the occupation numbers alone. The choice of statistics is therefore not a negligible refinement at the temperatures relevant to the QGP. This motivates our use of Fermi-Dirac statistics throughout the magnetized-medium quantities of Sec.~\ref{subsec:magnetic_corrections}, rather than the Boltzmann approximation of Ref.~\cite{Cleymans:1992je}.

The right panel of Fig.~\ref{fig:bm-fd-comparison} compares the Fermi--Dirac bremsstrahlung rate with LPM suppression, again evaluated with $L_C=1$, against its incoherent limit, obtained by dropping the LPM term $16q/(9\omega)$ from Eq.~\eqref{eq:vp_rate_221}, and against the $q\bar q$ annihilation rate at the same two temperatures. For the latter we use the lowest-order expression with massless Fermi--Dirac quarks~\cite{McLerran:1984ay,Kajantie:1986dh},
\begin{multline}
    \frac{dN^{q\bar q}}{d^4x\,dM} = \frac{2N_c\,\alpha^2}{3\pi^3}\Big(\sum_f e_f^2\Big) M T \\ \int_0^\infty dk\; \,\frac{k}{k_0}\, f_B(k_0)\,\ln\!\left[\frac{\cosh\!\big(\tfrac{k_0+k}{4T}\big)}{\cosh\!\big(\tfrac{k_0-k}{4T}\big)}\right],
\label{eq:anni}
\end{multline}
where $k_0=\sqrt{k^2+M^2}$, $f_B(k_0)=1/(e^{k_0/T}-1)$ and $\sum_f e_f^2=5/9$ for $u,d$ quarks.
At $M=0.05$~GeV the LPM effect reduces the bremsstrahlung rate by a factor $\approx13$ at $T=0.15$~GeV and $\approx48$ at $T=0.30$~GeV. The suppression fades with increasing $M$: at $M=1$~GeV the LPM rate is within $\approx1\%$ ($T=0.15$~GeV) and $\approx12\%$ ($T=0.30$~GeV) of the incoherent one. With LPM included, annihilation overtakes bremsstrahlung at $M\approx0.25$~GeV for $T=0.15$~GeV and at $M\approx0.50$~GeV for $T=0.30$~GeV, so bremsstrahlung dominates only in the low-mass region, as stated in Sec.~\ref{subsec:thermal_recap}. The incoherent bremsstrahlung rate, i.e.\ without the LPM term, would instead exceed annihilation by approximately three to four orders of magnitude at $M=0.05$~GeV, which shows that the LPM suppression is essential for a realistic estimate of the low-mass yield.

\begin{figure*}[tbh]
    \centering
    \includegraphics[width=\linewidth]{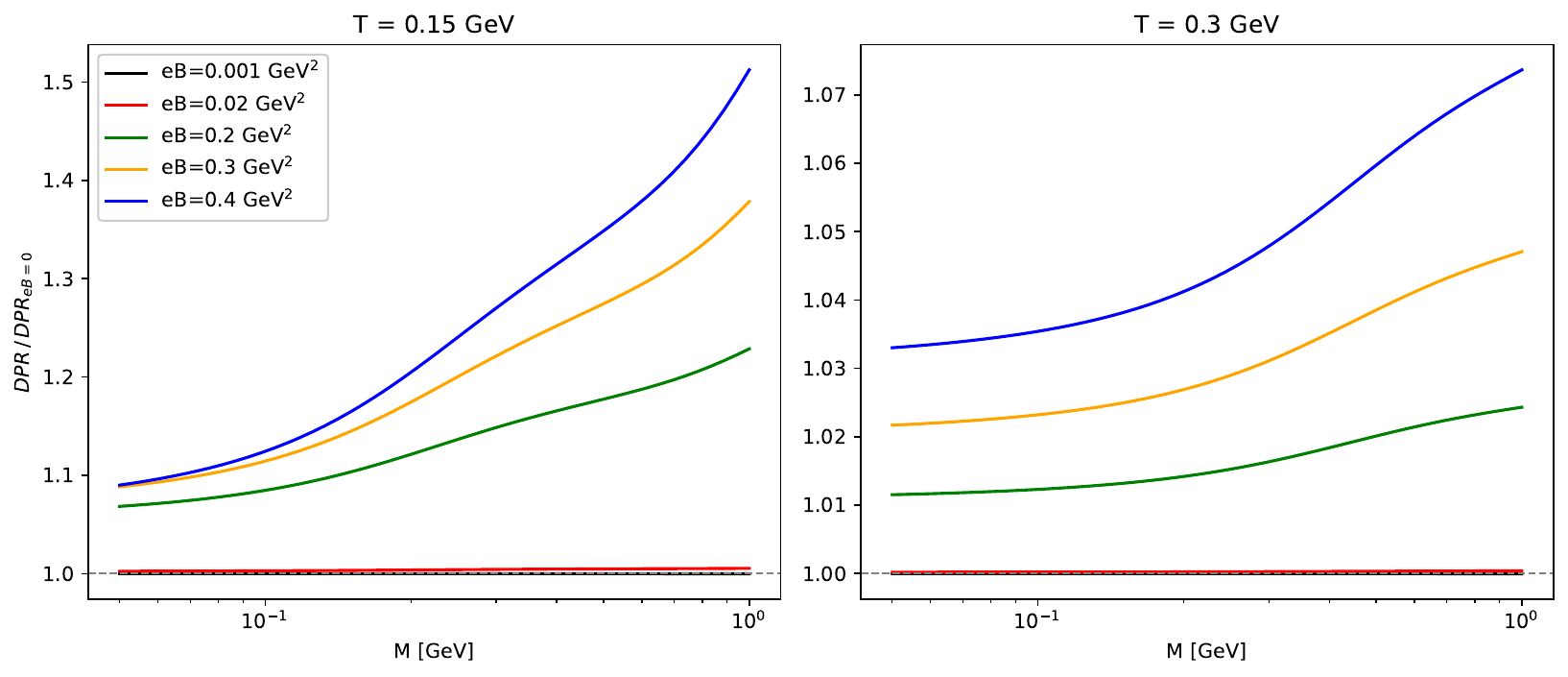}
    \caption{Ratio of the dilepton production rate $DPR = dN/d^4x\,dM$ to its $eB=0$ value as a function of the virtual photon invariant mass $M$, for different values of the external magnetic field, at $T=0.15$~GeV (left) and $T=0.30$~GeV (right).}
    \label{fig:dilepton-rates}
\end{figure*}

\subsection{The LPM parameter in a magnetized medium}
\label{subsec:q-vs-p1}

Figure~\ref{fig:q-vs-p1} shows $q(T,eB,p_1)$ of Eq.~\eqref{eq:q_magnetised} as a function of the quark momentum $p_1$, for the field strengths and temperatures of Table~\ref{tab:kmax-convergence}. At fixed $T$ and $eB$, $q$ is non-monotonic in $p_1$: it rises steeply just above the lower cutoff $p_1=m_D(T,eB)$, where the Coulomb logarithm $\ln(p_1/m_D)$ is small, and falls off at large $p_1$ once the $1/p_1^2$ prefactor in Eq.~\eqref{eq:q_magnetised} dominates. The weakest fields, $eB=0.001$ and $0.02~\mathrm{GeV}^2$, sit indistinguishably on top of the field-free $q_0(T,p_1)$ curve (dashed), consistent with the $n(T,eB)/n_0(T)\to1$ and $m_D(T,eB)/m_{D0}(T)\to1$ limits established in Sec.~\ref{subsec:kmax-convergence}. As $eB$ increases toward $0.4~\mathrm{GeV}^2$, the curve is enhanced across the full $p_1$ range shown, with the separation from the field-free result visibly larger at $T=0.15$~GeV than at $T=0.30$~GeV -- reflecting the same $eB/T^2$ scaling that controls the enhancement of $n(T,eB)$ itself in Table~\ref{tab:kmax-convergence}.

\subsection{Dilepton production rate}
\label{subsec:dilepton-rates}

Figure~\ref{fig:dilepton-rates} shows the ratio of the dilepton production rate $dN/d^4x\,dM$ through virtual bremsstrahlung in the presence of an external magnetic field to the $eB=0$ rate, as a function of the invariant mass $M$, for the same field strengths and temperatures as Fig.~\ref{fig:q-vs-p1}. This shows the isolated magnetic-field effect from the overall thermal falloff with $M$. As in Fig.~\ref{fig:q-vs-p1}, $eB=0.001$ and $0.02~\mathrm{GeV}^2$ stay within $1\%$ of unity across the full $M$ range at both temperatures, while $eB=0.2$--$0.4~\mathrm{GeV}^2$ produce a visible enhancement that grows with $M$: e.g. at $T=0.15$~GeV the strongest field considered, $eB=0.4~\mathrm{GeV}^2$, enhances the rate by $\approx9\%$ at $M=0.05$~GeV, rising to $\approx51\%$ at $M=1$~GeV; at $T=0.30$~GeV the same field gives a more modest $\approx3\%$ to $\approx7\%$ enhancement over the same range. The intermediate field $eB=0.3~\mathrm{GeV}^2$ tracks consistently between the $eB=0.2$ and $eB=0.4~\mathrm{GeV}^2$ curves at both temperatures. The enhancement follows from the growth of $n(T,eB)/n_0(T)$ with $eB$ (Table~\ref{tab:kmax-convergence}), which raises $q$ and hence the coherent term in Eq.~\eqref{eq:vp_rate_221}, partly offset by the increase of $m_D(T,eB)$ in the logarithm. The substantially larger effect at the lower temperature is again consistent with the field strength entering the magnetized-medium quantities through the dimensionless ratio $eB/T^2$, which is a factor of four larger at $T=0.15$~GeV than at $T=0.30$~GeV for the same $eB$.

This enhancement should be read against the scale hierarchy $eB\lesssim M^2$ of Eq.~\eqref{eq:scale_ratio}, which justifies the field-free vertex phase space. Table~\ref{tab:scale-check} lists $eB/M^2$ across the plotted range: the enhancement is largest at larger $M$, where the hierarchy is best satisfied, while at small $M$ the stronger-field curves indicate the trend and a fully consistent treatment there would require incorporating the magnetic field non-perturbatively in the production vertex, which we leave for future work.

\begin{table}[h]
\centering
\caption{$eB/M^2$ across the plotted $M$ range (in units of GeV), for the field strengths
of Fig.~\ref{fig:dilepton-rates}. The field-free vertex hierarchy of
Eq.~\eqref{eq:scale_ratio} requires this ratio $\lesssim 1$.}
\label{tab:scale-check}
\begin{tabular}{cccccccc}
\hline\hline
$eB$ [GeV$^2$] & $M=0.05$ & $0.1$ & $0.2$ & $0.3$ & $0.5$ & $0.7$ & $1.0$ \\
\hline
0.001 & 0.400 & 0.100 & 0.025 & 0.011 & 0.004 & 0.002 & 0.001 \\
0.02 & 8.00  & 2.00  & 0.50 & 0.222 & 0.080 & 0.041 & 0.020 \\
0.2  & 80.0  & 20.0  & 5.00 & 2.222 & 0.800 & 0.408 & 0.200 \\
0.3  & 120.0 & 30.0  & 7.50 & 3.333 & 1.200 & 0.612 & 0.300 \\
0.4  & 160.0 & 40.0  & 10.0 & 4.444 & 1.600 & 0.816 & 0.400 \\
\hline\hline
\end{tabular}
\end{table}

%% ============================================================
\section{Summary and Outlook}
\label{sec:conclusions}
%% ============================================================

Our treatment captures the leading magnetic-field imprint on the LPM-resummed dilepton invariant-mass spectrum through the field dependence of the equilibrium screening mass $m_D(T,eB)$ and quark density $n(T,eB)$, within a field-free phase-space approximation for the dilepton production vertex, valid for $eB \lesssim M^2$. We have argued that the off-equilibrium, field-driven correction to the quark distribution function, while nonzero in the fully differential rate, vanishes identically upon angular integration and therefore does not contribute to $dN/d^4x\,dM$ at linear order in the field, consistent with the numerical findings of Ref.~\cite{Gao2025} for quark-antiquark annihilation. The LPM mechanism thus carries a nonzero, computable magnetic-field signature in $dN/d^4x\,dM$, arising from the equilibrium medium quantities and independent of the vanishing linear-in-$\mathbf{E}$ correction. Quantitatively, for $eB=0.4~\mathrm{GeV}^2$ the bremsstrahlung rate is enhanced by $\approx9\%$ ($M=0.05$~GeV) to $\approx51\%$ ($M=1$~GeV) at $T=0.15$~GeV, and by $\approx3\%$ to $\approx7\%$ at $T=0.30$~GeV, while fields $eB\le0.02~\mathrm{GeV}^2$ change the rate by less than $1\%$. The LPM suppression itself (at constant $L_C$) reduces the field-free rate by factors $\approx13$ and $\approx48$ at $M=0.05$~GeV for $T=0.15$ and $0.30$~GeV, respectively, so that annihilation dominates above $M\approx0.25$ and $0.50$~GeV. These results are obtained for quark scatterers only, consistently with the quark-only $n(T,eB)$ and $m_D(T,eB)$ used throughout.

A fully field-theoretic treatment, incorporating the Schwinger proper-time representation of the quark propagator directly into the photon self-energy, would be required to capture Landau-level threshold structure and anisotropic propagation effects beyond the present Debye screening mass and number density mediated approximation, and is left to future work. Extending the present formalism to angle-differential observables such as the dilepton elliptic flow $v_2(M)$, where the off-equilibrium correction survives the relevant angular projection and is expected to dominate over the equilibrium channel identified here, is a natural next step.

\begin{acknowledgements}
A.~B. acknowledges support from the ULAM fellowship program of the Polish National Agency for Academic Exchange (NAWA), No.~BNI/ULM/2024/1/00193. The work of K.~R. and C.~S. is supported by the National Science Centre (NCN), Poland, under OPUS Grant No.~2022/45/B/ST2/01527. K.~R. acknowledges the support of the Polish Ministry of Science and Higher Education. C.~S. acknowledges the support of the World Premier International Research Center Initiative (WPI) under MEXT, Japan.
\end{acknowledgements}

\bibliography{LPM}

\end{document}